\documentclass[prl,preprint,floatfix]{revtex4}

\usepackage{graphicx}

\begin{document}

	\title{Local observation of linear-$T$ superfluid density \\and anomalous vortex dynamics in URu$_2$Si$_2$}

	\author{Yusuke Iguchi$^{1,2,4}$, Irene P. Zhang$^{1,2}$, Eric D. Bauer$^{3}$, Filip Ronning$^{3}$, John R. Kirtley$^{4}$, and Kathryn A. Moler$^{1,2,4}$} 
	\affiliation{$^1$Department of Applied Physics, Stanford University, Stanford, California 94305, USA\\
		$^2$Stanford Institute for Materials and Energy Sciences, SLAC National Accelerator Laboratory, 2575 Sand Hill Road, Menlo Park, California 94025, USA\\
		$^3$Los Alamos National Laboratory, Los Alamos, New Mexico 87545, USA\\
		$^4$Geballe Laboratory for Advanced Materials, Stanford University, Stanford, California 94305, USA}
	
	%\date{}
	
	\begin{abstract}
	\bf{
	    The heavy fermion superconductor URu$_2$Si$_2$ is a candidate for chiral, time-reversal symmetry-breaking superconductivity with a nodal gap structure. Here, we microscopically visualized superconductivity and spatially inhomogeneous ferromagnetism in URu$_2$Si$_2$. We observed linear-$T$ superfluid density, consistent with {\it d}-wave pairing symmetries including chiral {\it d}-wave, but did not observe the spontaneous magnetization expected for chiral {\it d}-wave. Local vortex pinning potentials had either four- or two-fold rotational symmetries with various orientations at different locations. Taken together, these data support a nodal gap structure in URu$_2$Si$_2$ and suggest that chirality either is not present or does not lead to detectable spontaneous magnetization.}
	    %Abstract length is 630[550] characters with spaces [no spaces]
	    %Abstract length <= 600 characters for PRL
	    
		%The heavy fermion superconductor URu$_2$Si$_2$ is a candidate for chiral, time-reversal symmetry-breaking superconductivity with a nodal gap structure. However, the signatures of chirality and nodes appear in some crystals but not in others, raising questions about the true nature of the superconducting state. Here, in order to measure spontaneous magnetism, the superfluid density and local pinning potentials, we imaged the zero-field magnetic flux, the low-field diamagnetic response, and the dynamics of isolated vortices in URu$_2$Si$_2$ with micron-scale spatial resolution, using scanning superconducting quantum interference device microscopy. We observed superconductivity and spatially inhomogeneous ferromagnetism. At non-ferromagnetic area, we observed linear-$T$ superfluid density, consistent with {\it d}-wave pairing symmetries including chiral {\it d}-wave forms of $k_z(k_x+ik_y)$, but no spontaneous magnetization expected for chiral superconductivity. Local vortex pinning potentials had either four- or two-fold rotational symmetries with various orientations at different locations. Taken together, these data provide evidence for a nodal gap structure and robust superconductivity coexisting on micron scales with inhomogeneous ferromagnetism and the upper limit size of chiral domains in URu$_2$Si$_2$.}
	\end{abstract}
	%\pacs{}
	\maketitle
	
	%Intro
	The heavy fermion superconductor URu$_2$Si$_2$ has been extensively studied to reveal the order parameters of the enigmatic hidden order (HO) phase (with critical temperature $T_{\mbox{\scriptsize HO}} =$ 17.5 K) and the coexisting unconventional superconducting (SC) phase (with critical temperature $T_{\mbox{\scriptsize c}} =$ 1.5 K)\cite{Mydosh2011,Shibauchi2014}. In the HO phase of URu$_2$Si$_2$, the small size of the (possibly extrinsic) magnetic moment previously detected by neutron scattering measurements is inconsistent with the magnitude of the large heat capacity anomaly at the transition\cite{Broholm1987}. Recent, though controversial, measurements of the magnetic torque\cite{Okazaki2011s}, the cyclotron resonance\cite{Tonegawa2012}, and the elastoresistivity\cite{Riggs2014} imply that HO phase has an electronic nematic character, reducing the four-fold rotational symmetry of the tetragonal lattice structure to two-fold rotational symmetry. Although the crystal lattice is also weakly forced to transform into an orthorhombic symmetry in ultra-pure samples\cite{Tonegawa2014}, the structural phase transition temperature differs from $T_{\mbox{\scriptsize HO}}$ at hydrostatic pressure\cite{Choi2018}. In response to these experiments, many theoretical models for the order parameter in the HO phase have been proposed, such as multipole orders\cite{Kiss2005,Haule2009,Kusunose2011,Cricchio2009,Ikeda2012}, but this order parameter is still not well understood. Further, although the HO phase coexists with the SC phase, it is unclear whether and how these phases are correlated.

    Recent studies suggest that the SC order parameter of URu$_2$Si$_2$ most likely possesses a chiral {\it d}-wave symmetry\cite{Shibauchi2014}. Knight shift measurements\cite{Knetsch1993,Hattori2018} and upper critical field $H_{c2}$ measurements\cite{Brison1995} both suggest a spin singlet state. Further, nodal gap structures were indicated by point contact spectroscopy measurements\cite{Hasselbach1993_46}, electron specific heat\cite{Hasselbach1993_47,Yano2008,Kittaka2016}, NMR relaxation rate \cite{Kohori1996}, and thermal transport measurements\cite{Kasahara2007,Kasahara2009}. Thermal conductivity measurements suggested the presence of a horizontal line node $L_{\mbox{\scriptsize H}}$ in the light hole band and point nodes $2P$ at the north and south poles in the heavy electron band\cite{Kasahara2007,Kasahara2009}. Similarly one electronic specific heat measurement also suggested the presence of point nodes in the heavy electron band\cite{Yano2008}, but a recent experiment detected the line node $L_{\mbox{\scriptsize H}}$ in the heavy electron band\cite{Kittaka2016}. In addition, a largely enhanced Nernst effect has been observed above $T_{\mbox{\scriptsize c}}$, which was explained as an effect of chiral phase fluctuations\cite{Yamashita2015}. Spontaneous time-reversal symmetry breaking in the SC phase was revealed by a polar Kerr effect measurement\cite{Schemm2015}. In addition, ferromagnetic (FM) impurity phases also have been implicated by nonlocal magnetization measurements\cite{Uemura2005,Amitsuka2007} and one polar Kerr effect measurement\cite{Schemm2015}.

	Here we sought to clarify the local time-reversal symmetry, the correlation between the HO and the SC phases, and the SC pairing symmetry in URu$_2$Si$_2$ by examining local magnetic fluxes and local superfluid responses.
	We used a local magnetic probe microscope called a scanning Superconducting QUantum Interference Device (SQUID) microscope (Fig. 1a). Scanning SQUID microscopy (SSM) has been used to scan the local magnetization of candidate chiral superconductors, which provided limits on the size of chiral domains by comparing experimental noise with theoretically expected magnetization\cite{Cliff2010}, and to image the magnetism in the superconducting ferromagnet UCoGe\cite{Hykel2014}. SSM also revealed stripe anomalies in the susceptibility along twin boundaries  near $T_{\mbox{\scriptsize c}}$ in iron-based superconductors\cite{Beena2010,Irene} and a copper oxide superconductor\cite{Logan2019}. Recently, local anisotropic vortex dynamics along twin boundaries were observed via SSM in a nematic superconductor FeSe\cite{Irene}. In addition, the local London penetration depth $\lambda$ can be estimated by the scanning SQUID height dependence of the local susceptibility\cite{kirtleyprb2012}. Therefore, SSM provides information of spontaneous magnetism, rotational symmetry of lattice structures, and superconducting gap structures in situ.

	%Method
	We used SSM to locally obtain the dc magnetic flux and ac susceptibility on the cleaved c-plane of single crystals of URu$_2$Si$_2$ (Figs. 1b,c) at temperatures varying from 0.3 K to 18 K using a Bluefors LD dilution refrigerator\cite{supple}. Bulk single crystals of URu$_2$Si$_2$ were grown via the Czochralski technique and electro-refined to improve purity\cite{Schemm2015}. Our scanning SQUID susceptometer had two pickup loop (PL) and field coil (FC) pairs (Fig. 1a) configured with a gradiometric structure\cite{kirtleyrsi2016}. The PL provides the local dc magnetic flux $\Phi$ in units of the flux quantum $\Phi_0=h/2e$, where $h$ is the Planck constant and $e$ is the elementary charge. The PL also detects the ac magnetic flux $\Phi^{ac}$ in response to the ac magnetic field $He^{i\omega t}$, which was produced by an ac current of $|I^{ac}| =$ 3 mA at 150 Hz through the FC, using an SR830 Lock-in-Amplifier. Here we report the local ac susceptibility as $\chi=\Phi^{ac}/|I^{ac}|$ in units of $\Phi_0/A$ and the local flux $\Phi$ as $\phi=\Phi/\Phi_0$.

    \begin{figure}[hb]
		\begin{center}
			\includegraphics*[width=12cm]{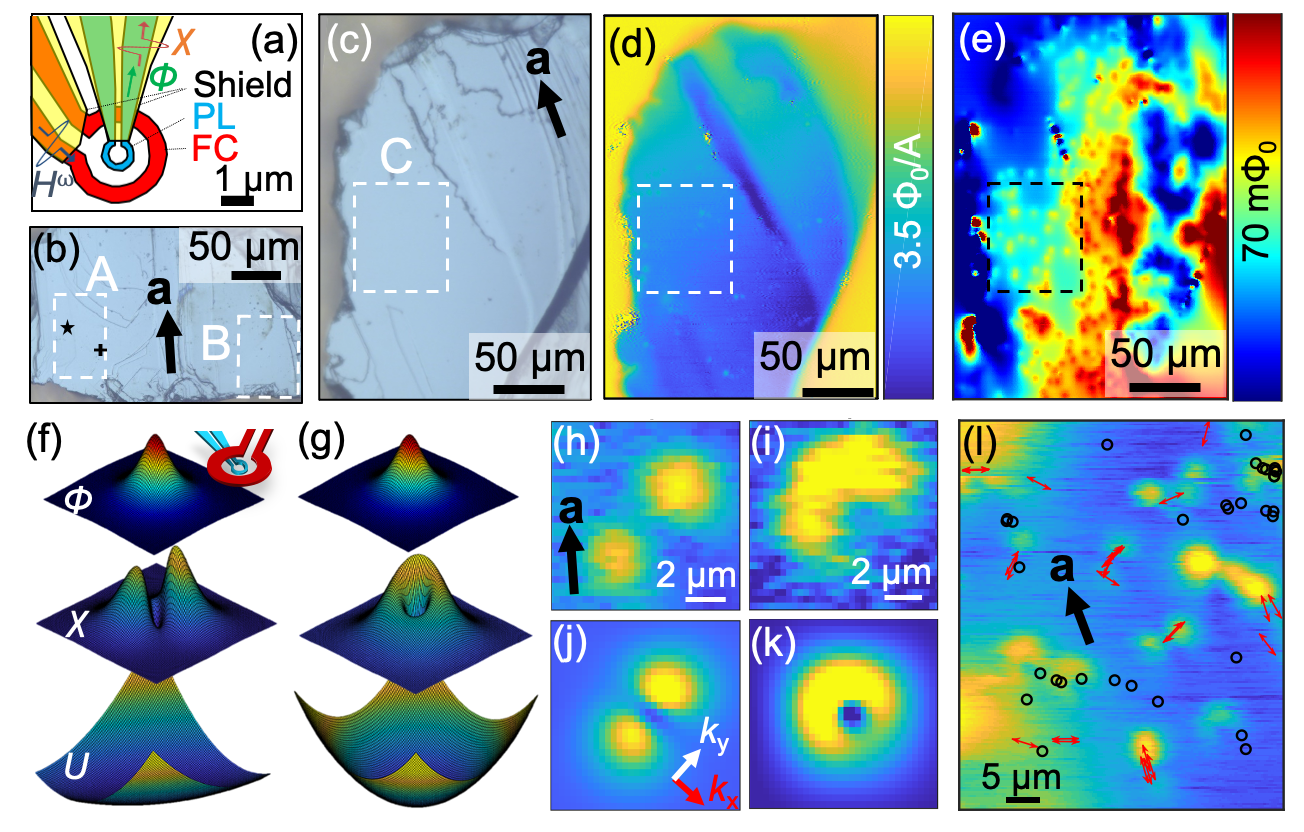}
			\caption{ SSM imaged inhomogeneous magnetic fluxes, superfluid response, and anomalous isolated vortex dynamics. (a) PL and FC of our SQUID susceptometer are covered with superconducting shields except for the loop area to detect local magnetic flux. (b),(c) Optical images of (b) sample 1 and (c) sample 2. We examined scanning SQUID measurements at flat regions A, B and C. (d),(e) In sample 2, (d) $\chi$ and (e) $\phi$ values acquired at $T =$ 0.5 K. (f),(g) SSM directly images isolated vortex dynamics. Schematics of SSM measuring $\phi$ and $\chi$ over an isolated vortex, where $U$ is (f) anisotropic or (g) isotropic. (h),(i) $\chi$ values over an isolated vortex acquired at (h) the star mark of region A at $T =$ 1.0 K, and (i) the plus mark of region A at $T =$ 1.2 K. (j),(k) Simulated $\chi$ values obtained by using (j) $k_x =$ 107.7 nN/m, $k_y =$ 19.9 nN/m, and (k) $k_x = k_y =$ 17.6 nN/m to capture (h) and (i), respectively. (l) Local rotational symmetry of $U$ varies randomly on a microscopic scale. $\chi$ values in region C at $T =$ 1.2 K. Black open circles and red double ended arrows indicate the isotropic and anisotropic vortex dynamics, respectively, which were observed at 1 K in different cooling cycles. The full scale variation in $\chi$ in images of (h)-(l) is 0.7 $\Phi_0$/A. Black single ended arrows indicate {\it a} axis.}
		\end{center}
	\end{figure}

	%Results 1
	We cooled samples from $T =$ 5 K to $T =$ 0.5 K with a dc magnetic field to produce the vortices. Then we observed inhomogeneity in the local susceptibility (Fig. 1d, sample 2) and the local magnetic flux (Fig. 1e, sample 2). Strong diamagnetic susceptibility due to the Meissner effect was only detected inside the sample(Fig. 1d); the inhomogeneity of $\chi$ mainly results from surface roughness(Fig. 1d). In contrast to the almost homogeneous Meissner effect observed on the whole sample, we detected FM domains on the right side of the sample, and many vortices on the left side(Fig. 1e, sample 2).

	We also observed local vortex dynamics of sample 1(Figs. 1h,i) and of sample 2(Fig. 1l). Figures 1f and 1g schematically show the values of $\phi$ and $\chi$ expected for an isolated vortex if the vortex pinning potentials $U$ are anisotropic or isotropic, respectively. Local vortex pinning potentials can be inferred from scanning SQUID measurements of isolated vortex dynamics by modeling a simple quadratic pinning potential $U(\Delta x,\Delta y) = \frac{1}{2}\left( k_x \Delta x^2 + k_y \Delta y^2 \right)$, where $k_x$ and $k_y$ are the vortex pinning force constants and $\Delta x$ and $\Delta y$ are the displacement of the vortex center from the equilibrium point\cite{Irene}. Note that screening from the SC shields on the probe provide an additional asymmetry, which we reproduce in our numeric simulations. Thus, local ac susceptibility scans reveal the local rotational symmetry of pinning potentials. We observed two types of $\chi$ images around an isolated vortex in different locations of region A (Fig. 1h,i). The anisotropic data (Fig. 1h) look similar to the anisotropic vortex dynamics ($k_x \neq k_y$) observed by our similar measurement of SSM in FeSe\cite{Irene}, but on the other hand, the isotropic data (Fig. 1i) look similar to the isotropic vortex dynamics ($k_x = k_y$) numerically simulated in \cite{Irene}. Our simulations reproduced the experimental data (Fig. 1j, anisotropic; Fig. 1k, isotropic)\cite{supple}. Our measurements and simulations revealed that vortex pinning potentials had four-fold or two-fold rotational symmetries at different locations in the same sample on a microscopic scale(Figs. 1h-l).

	%Discussion 1
	Two types of vortex dynamics, anisotropic (Fig. 1h) and isotropic (Fig. 1i), were observed with various orientations at different locations of sample 1. The observed vortex pinning positions were not ordered. The observed vortex responses to an applied force are modeled by simulations with isotropic pinning potentials (Fig. 1j) and two-fold rotationally symmetric pinning potentials (Fig. 1k). One scenario, which causes locally isotropic and anisotropic vortex dynamics, is that local strain caused by local defects in the tetragonal crystal structure drives the anisotropic vortex pinning forces. This scenario is consistent with our data: the susceptibility images acquired near $T_{\mbox{\scriptsize c}}$ did not show the stripes along potential twin boundaries (Figs. 2b,c and Supplemental Figs. 1a,c\cite{supple}) that were previously reported in copper oxide\cite{Logan2019} and iron-based superconductors\cite{Beena2010,Irene}. The sample may have had a slightly orthorhombic crystal structures, but if so, its effect on the local vortex dynamics was so small that we could not detect it. Thus, we suspect that our observed anisotropic vortex pinning force may have been caused by local strain from point defects in our URu$_2$Si$_2$ samples.

	\begin{figure}[hb]
		\begin{center}
			\includegraphics*[width=15cm]{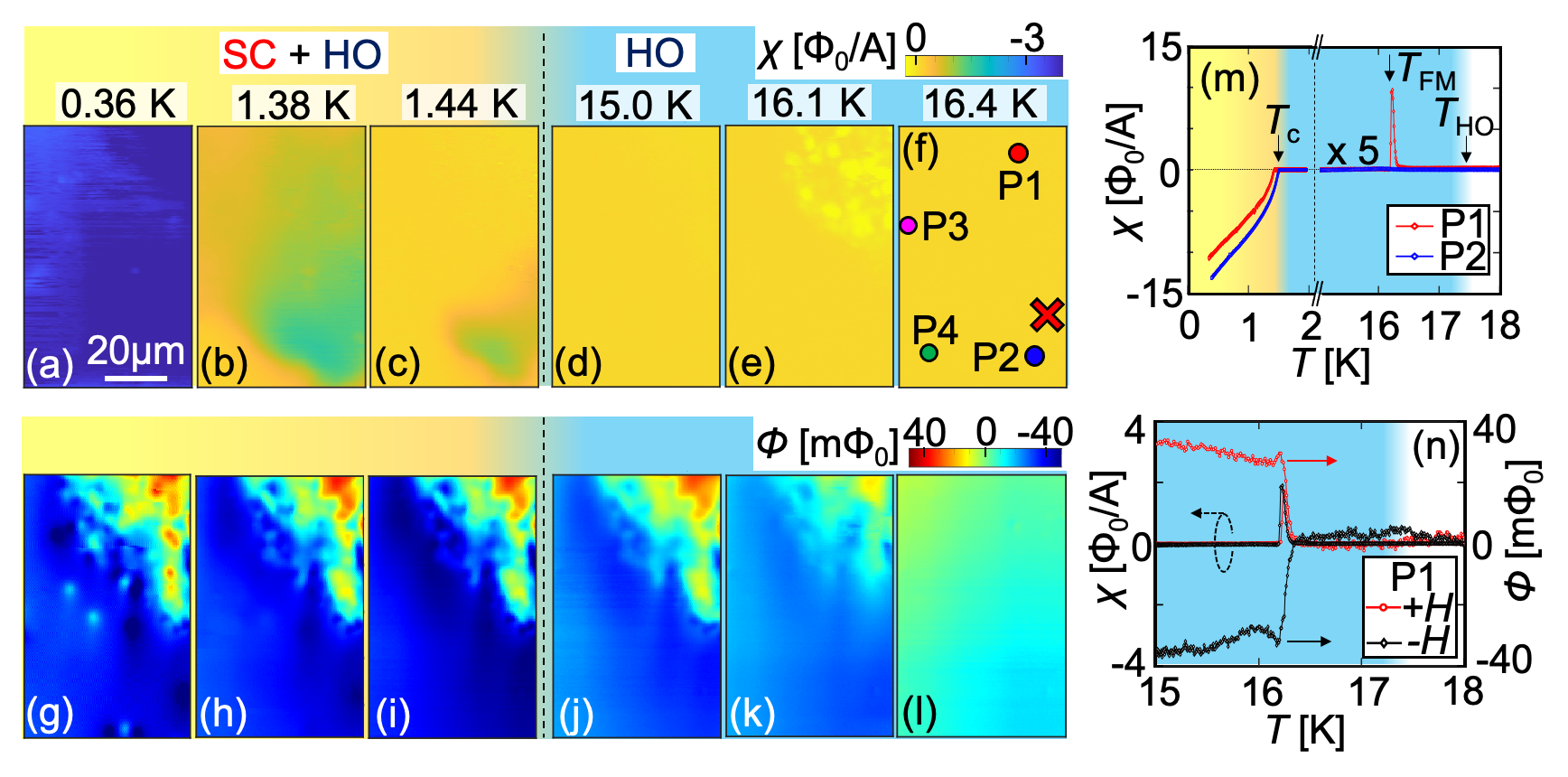}
			\caption{Superconductivity and ferromagnetism coexist locally. (a)-(l) Appearance of ferromagnetic domains and superconducting state visualized in (a)-(f) $\chi$ images and (g)-(l) $\phi$ images at $T =$ 0.36-16.4 K in region A of sample 1. (m) Ferromagnetism did not suppress superconductivity. $\chi$ above 15 K were plotted as 5 times experimental values to make these data easily viewable. (n) Ferromagnetic domain fields were oriented along the {\it c}-axis. $\chi$ and $\phi$ measured after field cooling, where $\mu_0H \sim$ 0.2 mT, at P1. $\mu_0$ is the permeability of free space.}
		\end{center}
	\end{figure}
	
	%Results 2
	Next, in order to examine correlations of the superconductivity, the ferromagnetism and the HO in URu$_2$Si$_2$, we determined the temperature dependence of $\chi$ and $\phi$ at region A of sample 1 (Figs. 2a-l). In the HO phase, $\chi$ and $\phi$ were homogeneous at $T =$ 16.4 K, but FM domains appeared in the upper right area below the HO transition. In this region an increase in the susceptibility $\chi$ was observed at 16.1 K (Figs. 2e,k), followed by a nearly constant magnetization $\phi$ below 15.0 K (Fig. 2d). In the coexisting SC + HO phase a negative $\chi$ appeared uniformly at 1.44 K (Fig. 2c). It is surprising that the FM domain continued to exist across $T_{\mbox{\scriptsize c}}$ and that it persisted even at 0.36 K, where the whole area showed strong diamagnetic $\chi$ (Fig. 2a). When we plotted the temperature dependence of $\chi$ at two specific points, the FM domain showed a sharp peak at 16.1 K (Fig. 2m). The direction of magnetic flux at the FM domain could be reversed by cooling the sample in a small applied dc magnetic field (Fig. 2n). There was no anomaly at $T_{\mbox{\scriptsize HO}}$ (Figs. 2m,n).
	
	%Discussion 2
	Although our investigations uncovered FM domains, we detected no spontaneous current propagating along sample edges or chiral domains. The expected spontaneous magnetization, which is carried by the chiral edge current, may be estimated by considering the orbital angular momentum of $\hbar l$ per Cooper pair, where $\hbar=h/2\pi$ and $l=$1 ({\it p}-wave), 2 ({\it d}-wave), or 3 ({\it f}-wave)\cite{Ishikawa1977,Tada2015,Nie2020}. This estimate neglects Meissner screening and surface effects, which will reduce the size of the effect. If the superconducting gap of URu$_2$Si$_2$ has chiral {\it d}-wave symmetry, the spontaneous magnetization $M_c$ is given by $e\hbar ln/4m^* \simeq$ 200 A/m, where $n$ is the carrier density and $m^*$ is the effective mass\cite{Tonegawa2013,Maple1986,Ohkuni1999,Okazaki2008,Kasahara2007,supple}. More careful calculations of the chiral edge current based on Bogoliubov-de Gennes analysis\cite{MatsumotoSigrist1999,Tada2015,Nie2020} showed that the signal is reduced by the Meissner screening current and surface effects, and also that the orbital angular momentum of the Cooper pair is suppressed due to multiple current modes\cite{Tada2015,Nie2020} and depairing effect\cite{Tada2015} for $l\geq2$ because the chiral edge current and the orbital angular momentum are not topologically protected properties. We also consider the possibility of random domains magnetized along the {\it c} axis including Meissner screening \cite{Bluhm2007}. For  large domains ($> \sim$10 $\mu$m), the scanning SQUID could resolve individual domain boundaries. The expected magnetic flux along the domain boundary for our experimental setup is estimated as $\sim$100 m$\Phi_0$ from the expected spontaneous magnetization of $M_c$ = 200 A/m and could be as low as 20 A/m [$\sim$10 m$\Phi_0$] after accounting for surface effects and multiple current modes\cite{supple}. For random domains of size of $L$ = 1 $\mu$m, the expected magnetic flux would have a random varying sign (depending on the local domain orientations) with a magnitude of about 4.1 m$\Phi_0$ [0.4 m$\Phi_0$] for $M_c$ = 200 A/m [20 A/m]\cite{supple}. The observed magnetic flux far from the FM domains was $\sim$0.5 m$\Phi_0$ in the PL, and its magnetic flux density was 3.5$\times$10$^{-6}$ T. For the expected spontaneous magnetization of $M_c$ = 200 A/m [20 A/m], we obtain a domain size limit of $L \leq$ 250 nm [1.1 $\mu$m], which is comparable to the size of our PL. It would be surprising to find domains that are so similar in size to the natural length scales of the superconductivity. Therefore, our measurements set an upper limit on spontaneous magnetization that suggests that  chiral superconductivity, if present, does not result in the estimated magnetic flux. However, some effects, such as surface effects or small domain structures, may have suppressed the spontaneous magnetization to levels below our sensor's detection limit.

	A FM signal was previously studied as an impurity effect\cite{Uemura2005,Amitsuka2007}. Amitsuka {\it et al}. used a commercial SQUID magnetometer to detect three FM phases in URu$_2$Si$_2$, $T_1^* = 120$ K, $T_2^* = 35$ K, and $T_3^* = 16.5$ K, which were all sample dependent \cite{Amitsuka2007}. Their neutron scattering results suggested that the magnetization in the $T_2^*$ phase was caused by the stacking faults of a $Q=(1,0,0)$ antiferromagnetic phase with a small moment. High-pressure scattering measurements revealed that the small-moment antiferromagnetic phase was spatially separated from the HO phase, and that the small moments originated from the small volume of the antiferromagnetic phase depending on the lattice ratio $c/a$\cite{Matsuda2001,Yokoyama2005}. Here, we clearly visualized that the $T_3^*$ phase makes FM domains but find no evidence of either $T_1^*$ or $T_2^*$ phases (Figs. 2j-l). The FM domains are spatially inhomogeneous, because positive peaks in susceptibility were only observed locally (Fig. 2e). In the SC phase, the FM domains coexist with superconductivity (Figs. 2a-c,g-i).	It is difficult to obtain a zero-field condition due to the long-range magnetic fields ($\sim$0.3 mT) produced by the FM domains (Supplemental Figs. S1 and S2\cite{supple}), but the FM domains surprisingly did not suppress the superconductivity of our samples. Thus, the superconductivity in URu$_2$Si$_2$ is robust against FM domains and disorders such as impurities and local strain, which are believed to be responsible for the FM $T_3^*$ phase. It remains possible, however, that the SC and the FM phases are spatially separated on a nanoscopic scale.

	\begin{figure}[hb]
		\begin{center}
			\includegraphics*[width=15cm]{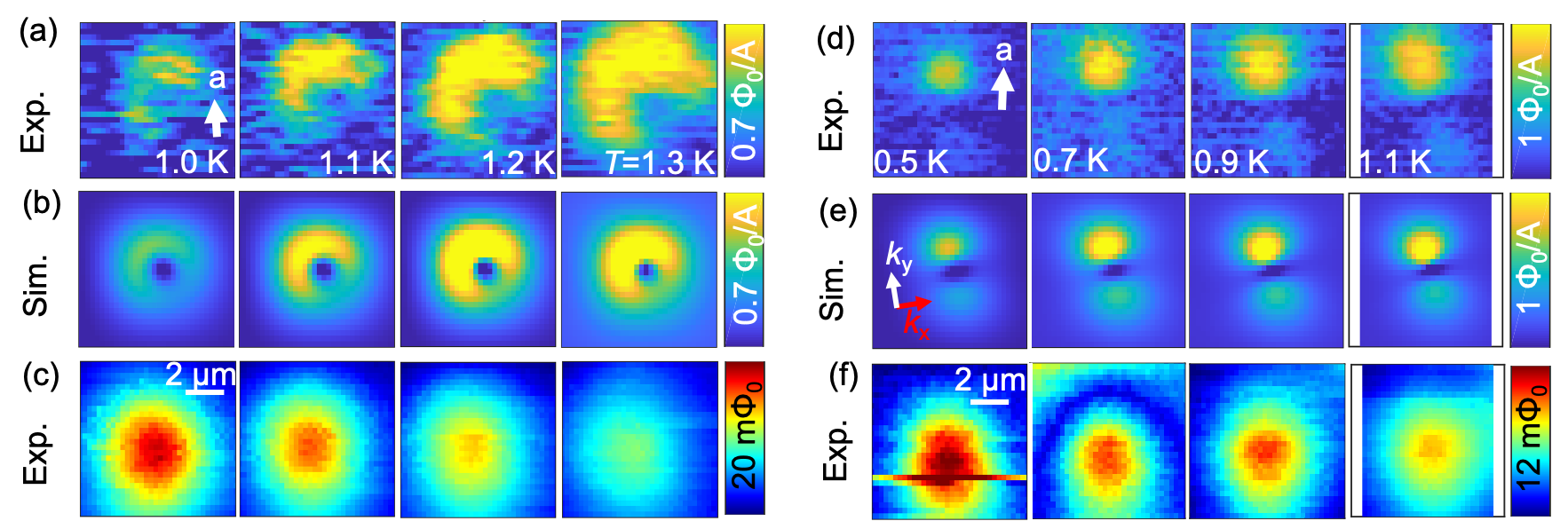}
			\caption{Isotropic or anisotropic vortex dynamics were enhanced near $T_{\mbox{\scriptsize c}}$, which are well explained by our simulation. (a)-(c) Temperature dependence of isotropic vortex dynamics in (a) experimental $\chi$ in region A of sample 1 (location denoted by a cross in Fig. 2f), and in (b) simulated $\chi$ with penetration depth obtained from the fitting of (c) the observed vortex field and pinning force constants ($k_x = k_y$). (d)-(f) Temperature dependence of anisotropic vortex dynamics in (d) experimental $\chi$ at region C of sample 2, and in (e) Simulated $\chi$ with the penetration depths from (f) the observed vortex field and various constants ($k_x \neq k_y$).}
		\end{center}
	\end{figure}
	
	%Results 3
	We experimentally obtained isotropic and anisotropic vortex dynamics (Figs. 3a,d) and vortex fields (Figs. 3c,f) and numerically simulated vortex dynamics (Figs. 3b,e). We obtained the local London penetration depth by fitting the magnetic flux from an isolated vortex\cite{kirtleyrsi2016} (Supplemental Fig. S3\cite{supple}). The simulations of the vortex dynamics have a systematically shorter spatial extent than experiments (Figs. 3 and Figs. S3\cite{supple}). We ignored these difference in the simulation, which may be caused by the error in the SQUID sensor height. By applying a $\chi^2$-test, we calculated the pinning force constants $k_x, k_y$ as 5-50 nN/m at $T = 1.3-1.0$ K for isotropic potentials, and as $k_x =$ 1-30 nN/m and $k_y/k_x =$ 5-10 at $T = 1.3-0.3$ K for anisotropic potentials, where $k_x > k_y$.  All obtained isotropic pinning force constants in regions A, B, and C had the same temperature dependence (Fig. 4a)\cite{supple}.
		
	%Discussion 3
	The temperature dependence of an isolated vortex pinning force has been discussed only in non-local measurements at small fields\cite{Ullmaier1975,Golosovsky1996}, but here we directly measured it. We use the hard core model \cite{Ullmaier1975}, where an isolated vortex cylinder core is pinned at a normal conducting small void, to fit the temperature dependence of an isolated vortex pinning force with constants $k \propto (1-(T/T_{\mbox{\scriptsize c}})^2)^m$, where $m$ depends on the dimensions of the small void. We obtain $m = 2$ from the best fit in Fig. 4(a), which indicates that our samples include small voids of roughly the same size as the coherence length\cite{Ullmaier1975}, $\sim $10 nm\cite{Amato1997}. The existence of nano-scaled voids supports our hypothesis that the local strain causes anisotropic and isotropic vortex dynamics at different locations of a URu$_2$Si$_2$ sample.
	
	\begin{figure}[hb]
		\begin{center}
			\includegraphics*[width=7cm]{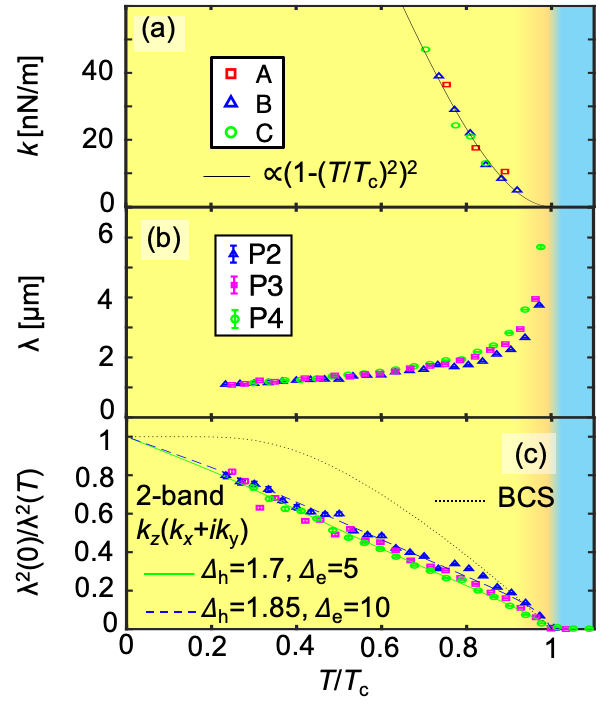}
			\caption{ (a) Vortex pinning force constants at three regions had the temperature dependence of $(1-(T/T_{\mbox{\scriptsize c}})^2)^2$. (b),(c) Superfluid density had a linear-{\it T} dependence at low temperature. (b) Temperature dependence of the penetration depth at three points. (c) Temperature dependence of normalized superfluid density from the penetration depths in (b), with $\lambda(0)$  =  1.0 $\mu$m. The dotted black line is the single s-wave gap BCS model for reference. The solid green line and dashed blue line are the two-band models for $k_z(k_x+ik_y)$ (light hole, heavy electron) with the indicated gap energies $\Delta_{\mbox{\scriptsize h,e}}$ in a unit of $k_{\mbox{\scriptsize B}}T_{\mbox{\scriptsize c}}$ to capture the experimental data.}
		\end{center}
	\end{figure}
	
	%Results 4
	The local London penetration depth $\lambda$ was determined by fitting the height dependence of susceptibility\cite{kirtleyprb2012} (Supplemental Fig. S4\cite{supple}). $\lambda$ at P2, P3, and P4 each saturated to approximately 1.0 $\pm$0.1 $\mu$m at zero temperature (Fig. 4b). These $\lambda$ values are quantitatively consistent with previous reports of $\lambda =$ 0.7-1.0 $\mu$m from measurements of muon spin relaxation\cite{Knetsch1993} and the estimate $\lambda = \sqrt{m^*/\mu_0ne^2} =$  1.1 $\mu$m\cite{supple}. We calculated the local superfluid density $n_s = \lambda^2(0)/\lambda^2(T)$ from the experimentally obtained $\lambda$ at P2, P3, and P4 (Fig. 4c). Here we determined $T_{\mbox{\scriptsize c}}$ at P2, P3, and P4 as 1.50, 1.41, and 1.34 K, respectively, by defining these as the temperatures where the superfluid density becomes almost zero. The superfluid density varied spatially near $T_{\mbox{\scriptsize c}}$, but all superfluid density values linearly increased as the normalized temperature decreased with temperature (Fig. 4c). 
	
	%Discussion 4
	The temperature dependence of the superfluid density in unconventional superconductors is estimated by the semi-classical approach with an anisotropic gap function\cite{Prozorov2006,supple}. Our results deviate from the numerically calculated superfluid density of the single band isotropic {\it s}-wave pairing symmetry model (BCS model)(Fig. 4c). The calculated curves for {\it d}-wave models are roughly consistent with our experimental results (Supplemental Fig. S5)\cite{sigristueda1991,supple}. However, they do not completely capture the behavior near $T_{\mbox{\scriptsize c}}$. In order to explain this difference, we used the two-band model $n_s = x n_h + (1-x) n_e$, where $x = 0.87$ is the ratio of the electron and hole mass, $n_h$ is the light hole band superfluid density, and $n_e$ is the heavy electron band superfluid density\cite{Prozorov2006,Okazaki2011}. Here we fit the experimental data with a model using chiral {\it d}-wave symmetry on the light hole and heavy electron bands with two free parameters of superconducting gaps $\Delta_{\mbox{\scriptsize h}}$ (hole band)and $\Delta_{\mbox{\scriptsize e}}$ (electron band)\cite{supple}, which well explain the experimental results (Fig. 4c, Supplemental Fig. S6a\cite{supple}). The fits to all {\it d}-wave symmetry two band models showed nearly identical results with different parameters (Supplemental Figs. S6b-e,S7a-b,S8\cite{supple}), but the two-band isotropic {\it s}-wave model's fitting results were markedly different from the experimental results (Supplemental Fig. S7c,S8)\cite{supple}. In particular, the values of $\Delta_{\mbox{\scriptsize h}}$ and $\Delta_{\mbox{\scriptsize e}}$, which were used in Fig. 4c, are almost same as values of $\Delta_{\mbox{\scriptsize h}}=1.6 k_{\mbox{\scriptsize B}}T_{\mbox{\scriptsize c}}$ and $\Delta_{\mbox{\scriptsize e}}=4 k_{\mbox{\scriptsize B}}T_{\mbox{\scriptsize c}}$ that were obtained from fits to the lower critical field $H_{c1}$ along the $a$ axis, which was measured with a Hall bar measurement\cite{Okazaki2011}. While we expect $n_s$ to exhibit the same temperature dependence as $H_{c1}$ along the $c$ axis, the Hall bar measurement report an anomalous kink structure at 1.2 K \cite{Okazaki2011}, which we did not observe in Fig. 4c. This difference may be a benefit of local measurements. For example, FM domains may affect $H_{c1}$ measurement only along the $c$ axis; here, FM domain fields had magnetic anisotropy along the $c$ axis (Figs. 1e, 2g-l, and Supplemental Fig. S1\cite{supple}) and the amplitude of a FM domain field is of the same order as the amplitude of $H_{c1}$ along the $c$ axis at 1.3 K\cite{Okazaki2011}. Thus, our model and experimental data clearly suggest the existence of nodal gap structures in URu$_2$Si$_2$, but it is difficult to distinguish distinct types of nodal gap structure by our data because the slope of linear-{\it T} superfluid density can be adjusted by the gap energies, which are free fit parameters in our model. 
	
	%Summary
	In summary, we have locally observed FM domains coexisting with superconductivity, local pinning potentials, and linear-$T$ superfluid densities in URu$_2$Si$_2$ on a microscopic scale. This superconductivity coexists robustly with inhomogeneous ferromagnetism on a micron scale, although we cannot tell if they coexist in the same physical volume on nanometer scales. Further, we detected no spontaneous magnetization associated with chiral domains in the SC phase. The obtained linear-$T$ superfluid density is well explained by {\it d}-wave models, but not by {\it s}-wave models. Taken together, these results provide new evidence for a nodal gap structure and robust superconductivity coexisting on micron scales with inhomogeneous ferromagnetism and place limits on the size of possible chiral domains in URu$_2$Si$_2$.
	
	%\end{document}
	
	%Acknowledgements
	The authors thank Ian R. Fisher and Steven A. Kivelson for fruitful discussion. This work was primarily supported by the Department of Energy, Office of Science, Basic Energy Sciences, Materials Sciences and Engineering Division, under Contract No. DE- AC02-76SF00515. Work at Los Alamos was performed under the auspices of the Department of Energy, Office of Science, Basic Energy Sciences, Materials Science and Engineering Division. Y.I. was supported by a JSPS Oversea Research Fellowship.

	%\section*{Acknowledgements}

	%\section*{Author contributions}
	%Y.I. carried out the scanning SQUID microscopy, the analyzed data, and the simulation of vortex dynamics. I.P.Z. and J.R.K. contributed to the simulation of vortex dynamics. E.D.B. and F.R. synthesized the crystals. K.A.M. supervised the project. Y.I. wrote the manuscript with input from all of the authors.

	\section*{Additional information} 
	The authors declare no competing financial interests. Correspondence and requests for materials should be addressed to Y. I. (yiguchi@stanford.edu)
	%Reprints and permissions information is available online at ().

	\newpage
	
	\setcounter{figure}{0}
	\renewcommand{\thefigure}{S\arabic{figure}}
	
	\begin{center}
		\Large
		%%%%%%%%%%%%%%%%%%%%%%%%%%%
		{Supplemental Material for \\\lq\lq Local observation of linear-$T$ superfluid density and anomalous vortex dynamics in URu$_2$Si$_2$ \rq\rq} \\by Iguchi $et$ $al.$
		%%%%%%%%%%%%%%%%%%%%%%%%%%%
	\end{center}
	
	\section*{1. Chiral superconductivity and ferromagnetism in URu$_2$Si$_2$ on scanning SQUID microscopy}

    We used scanning SQUID microscopy to locally obtain the dc magnetic flux and ac susceptibility of single crystals of URu$_2$Si$_2$ at temperatures varying from 0.3 K to 18 K using a Bluefors LD dilution refrigerator. The dimensions of samples 1 (Fig. 1b) and 2 (Fig. 1c) were $\sim0.4\times0.5\times0.05$ mm$^3$ and $\sim0.3\times0.7\times0.05$ mm$^3$, respectively. The widest surface on each sample was the cleaved c-plane. The pickup loops and field coils are covered with Nb superconducting shield layers (Fig. 1a); thus, only the magnetic flux going through the pickup loops in the SQUID loop was detected, and the external magnetic field was only applied by the FCs. The inner radius of the pickup loop was 0.3 $\mu$m, and the distance between the PL and the sample surface was $\sim$ 0.5 $\mu$m when the SQUID tip was touching the sample surface. 
	
	\begin{figure}[htb]
		\begin{center}
			\includegraphics*[width=11cm]{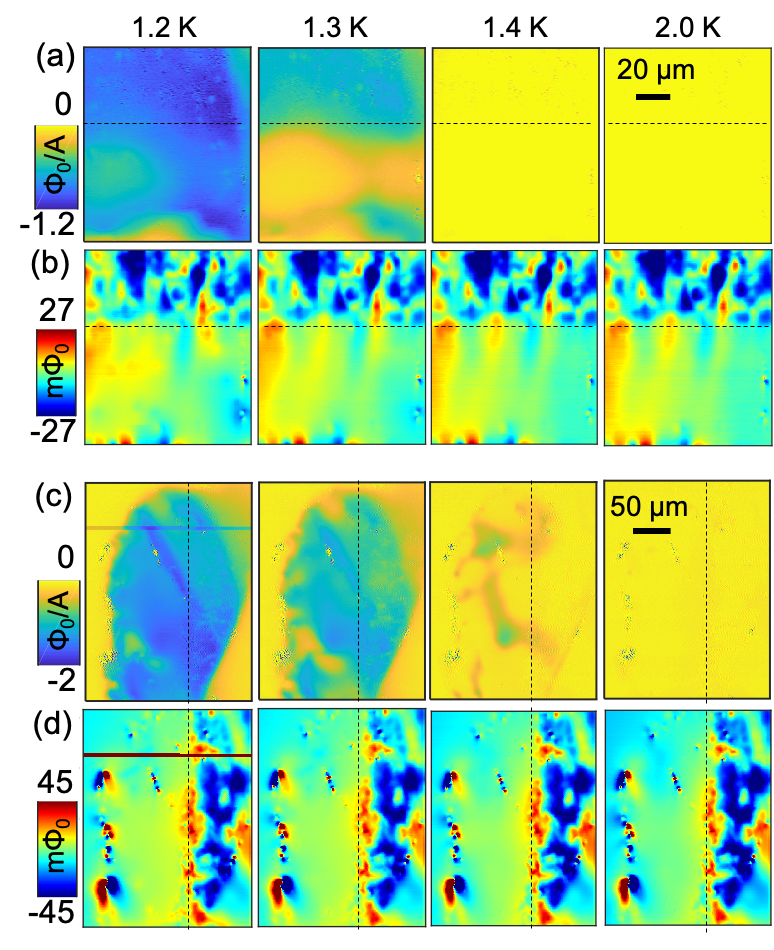}
			\caption{Superconductivity coexisted with ferromagnetism on a microscopic scale. (a),(b) Temperature dependence of (a) susceptibility and (b) magnetic flux images in region B of sample 1. (c),(d) Temperature dependence of (c) susceptibility and (d) magnetic flux images in sample 2. Dashed lines are guides for the eye to separate the FM domain from the paramagnetic domain.}
		\end{center}
	\end{figure}

	\begin{figure}[htb]
		\begin{center}
			\includegraphics*[width=8cm]{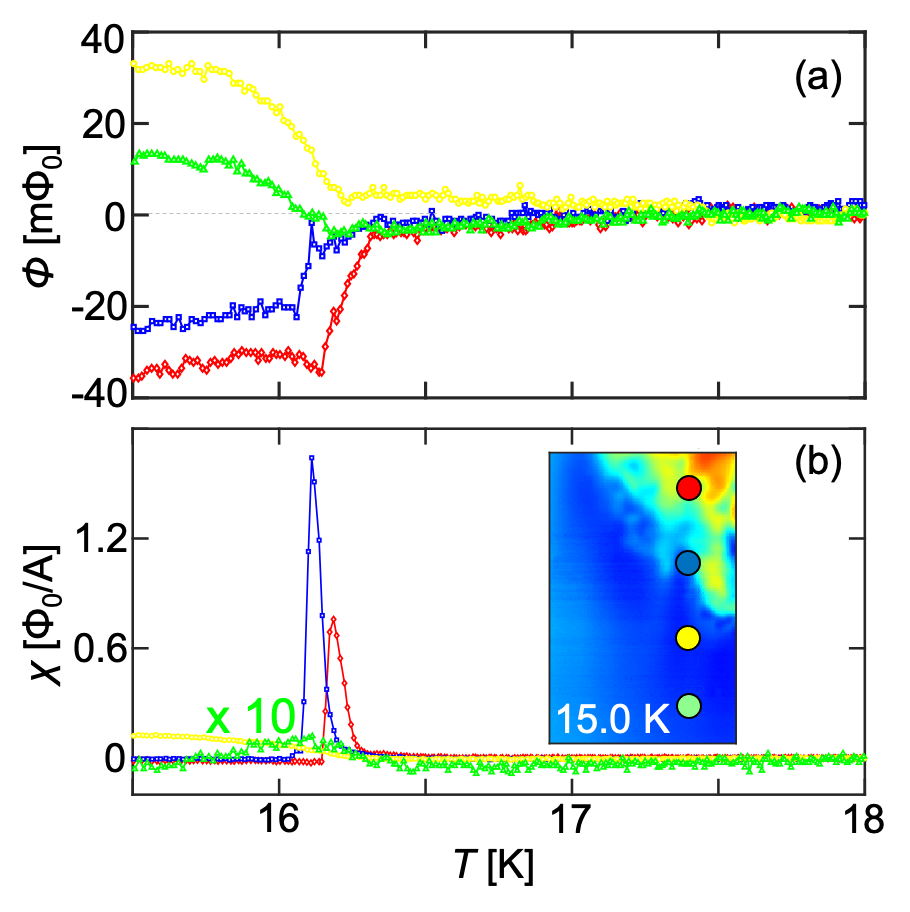}
			\caption{Ferromagnetic signals originated from the ferromagnetic domains. Temperature dependence of (a) magnetic flux and (b) susceptibility at the four positions. Inset indicates the positions on the magnetic flux image in region A at 15.0 K from Fig. 2b.}
		\end{center}
	\end{figure}
	
	The expected spontaneous magnetization, which is carried by the chiral edge current, may be overestimated by the orbital angular momentum of $\hbar l$ per Cooper pair, where $\hbar=h/2\pi$ and $l=$1 ({\it p}-wave), 2 ({\it d}-wave), or 3 ({\it f}-wave)\cite{Ishikawa1977,Tada2015,Nie2020}. For chiral {\it d}-wave symmetry, the spontaneous magnetization $M_c$ is given by
	\begin{equation}
	    M_c = \frac{1}{2V}|\int{r\times j dV}|=\frac{e\hbar ln}{4m^*},
	\end{equation}
	where current density $j=nep/2m^*$, $p$ is the angular momentum, $n$ is the carrier density, $V$ is a sample volume, and $m^*$ is the effective mass. By using $m_h^* \sim 13m_0$ (hole band), the averaged $m_e^* \sim 158m_0$ (electron band), and a carrier density of $2.6\times10^{28}$ m$^{-3}$ for both bands, we obtain $M_c = e\hbar n(1/m_h^* + 1/m_e^*)/2$ = 200 A/m, where $m_0$ is the free electron mass\cite{Tonegawa2013,Maple1986,Ohkuni1999,Okazaki2008,Kasahara2007}. 
	For large domains magnetized along the {\it c} axis where the domains are larger than the penetration depth in size, the expected magnetic field produced by $M_c sgn(x)$ oriented in the $z$-direction at height $z$ is given by\cite{Bluhm2007} 
	\begin{equation}
	    B_z(x,z) = \frac{\mu_0M_c}{\pi}\int{dk\frac{-ike^{ikx}e^{-|k|z}}{\sqrt{1/\lambda^2+k^2}\left( |k| + \sqrt{1/\lambda^2+k^2} \right)}}.
	\end{equation}
	For random domains magnetized along the {\it c} axis where the domains are comparable to or smaller than the penetration depth in size, the expected magnetic field near the domain boundary at $x=0$ produced by $M_c$ at height $z$ is given by 
	\begin{equation}
	    B_c^2 = \frac{15\pi V}{2}\frac{\lambda^3}{(z+\lambda)^6}\left(\frac{\mu_0 M_c}{4\pi}\right)^2, 
	\end{equation}
	where $V$ is the domain volume\cite{Bluhm2007}. For our experimental setup, we used $z$ = 0.5 $\mu$m and $\lambda$ = 1.0 $\mu$m to roughly estimate the limit of domain size of $L^3$.

    \section*{2. Isolated vortex dynamics on scanning SQUID microscopy}
    
    We observed local vortex dynamics of sample 1(Figs. 1h,i and Figs. 3a,c) and of sample 2(Fig. 1l and Figs. 3d,f). When we simulated vortex dynamics induced by ac magnetic field from the FC with these anisotropic and isotropic pinning potentials by calculating $\Phi^{ac} = \frac{d\phi}{dx}\Delta x + \frac{d\phi}{dy}\Delta y$, our simulations reproduced the experimental data (Fig. 1j, anisotropic; Fig. 1k, isotropic). The gradient of vortex field $\frac{d\phi}{dx}$ in our simulation was numerically calculated using the experimentally obtained London penetration depth\cite{kirtleyrsi2016}. The displacement of $\Delta x$ satisfies the equilibrium condition of forces $k_x\Delta x  = F_x$, where $F_x$ is the Lorentz force on an isolated vortex produced by an ac magnetic field\cite{Irene}. Here the four-fold rotational symmetry in the isotropic data (Fig. 1i) was broken, but this symmetry breaking was explained by the screening effect of the asymmetric superconducting shields in our simulations (Fig. 1k). It was also verified that the shield screening effect could not qualitatively change the apparent axis of anisotropic vortex dynamics in \cite{Irene}. The temperature dependence of pinning force constants at region A, B, and C were plotted by using the local $T_{\mbox{\scriptsize c}}$ = 1.48, 1.35, and 1.4 K were defined as the temperature where $\chi$ becomes zero (Fig. 4a).
    
    \begin{figure}[htb]
		\begin{center}
			\includegraphics*[width=10cm]{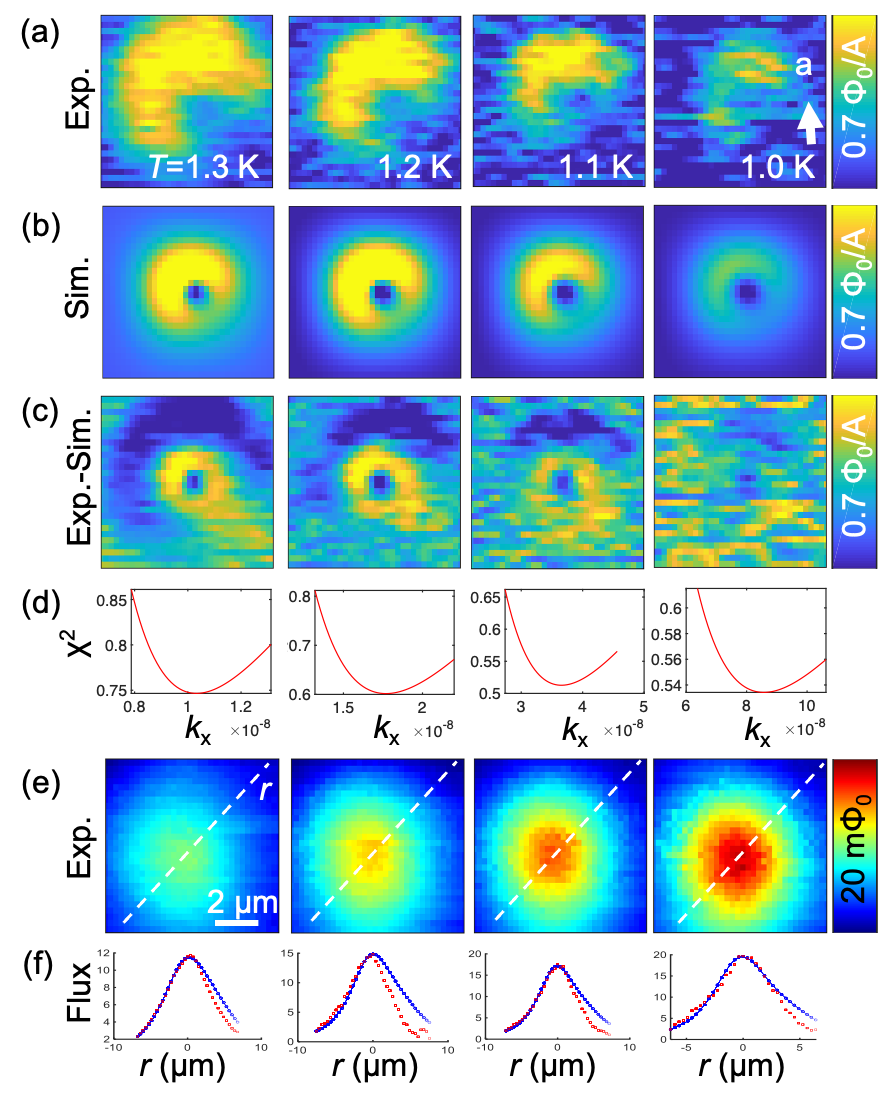}
			\caption{Isotropic vortex dynamics model captured the observed isotropic $\chi$ images. (a) Isotropic susceptibility images were acquired over (e) the isolated vortex at $T =$ 1.0, 1.1, 1.2, and 1.3 K in region A of sample 1 (location denoted by a cross in Fig. 2f). (b) Simulated susceptibility values were calculated by using (c),(d) $\chi^2$-test with penetration depths obtained from (f) the fitting of the experimental data and various spring constants ($k_x = k_y$) from Fig. 4a.}
		\end{center}
	\end{figure}

	%Discussion 1
	Two scenarios cause locally isotropic and anisotropic vortex dynamics. In the first scenario, local strain caused by local defects in tetragonal structure drives the anisotropic vortex pinning forces. As mentioned in the main text, this scenario is consistent with our data. In the second scenario, the twin boundaries drives the anisotropic vortex pinning forces, but the vortex pinning forces inside a large twin domain are isotropic. This second scenario is not consistent with our observation that the anisotropic vortex pinning forces were oriented along various directions at different locations (Fig. 1l). The anisotropy caused by twin boundaries and pinning locations in orthorhombic structure is expected to be along the tetragonal [100] or [010] directions\cite{Irene}. In addition, this second scenario is not consistent with the lack of strip anomaly in our susceptibility scans as mentioned above. However, an anomaly of susceptibility in URu$_2$Si$_2$ could be smaller than anomalies in iron-based superconductors, because URu$_2$Si$_2$ has an order of orthorhombicity that is two orders of magnitude smaller than that of BaFe$_2$As$_2$-based iron-pnictide superconductors\cite{Tonegawa2014}. Therefore, our observed anisotropic vortex pinning force could be caused by local strain.

    \section*{3. Superfluid density measurements and analysis on scanning SQUID microscopy}
    
    In order to obtain the local London penetration depth, we fitted the SQUID height dependence of local susceptibility (Supplemental Fig. S4) by using the theoretical equation (7) from Kirtley {\it et al.}\cite{kirtleyprb2012}, which includes the London penetration depth as a fitting parameter. Here the SQUID height $z$ included the thickness of the shield layer as 0.4 $\pm0.1$ $\mu$m, the error of which equally shifts the penetration depth by $\pm0.1$ $\mu$m. However the error in the superfluid density was sufficiently small to be ignored.
    
    \begin{figure}[b]
		\begin{center}
			\includegraphics*[width=8cm]{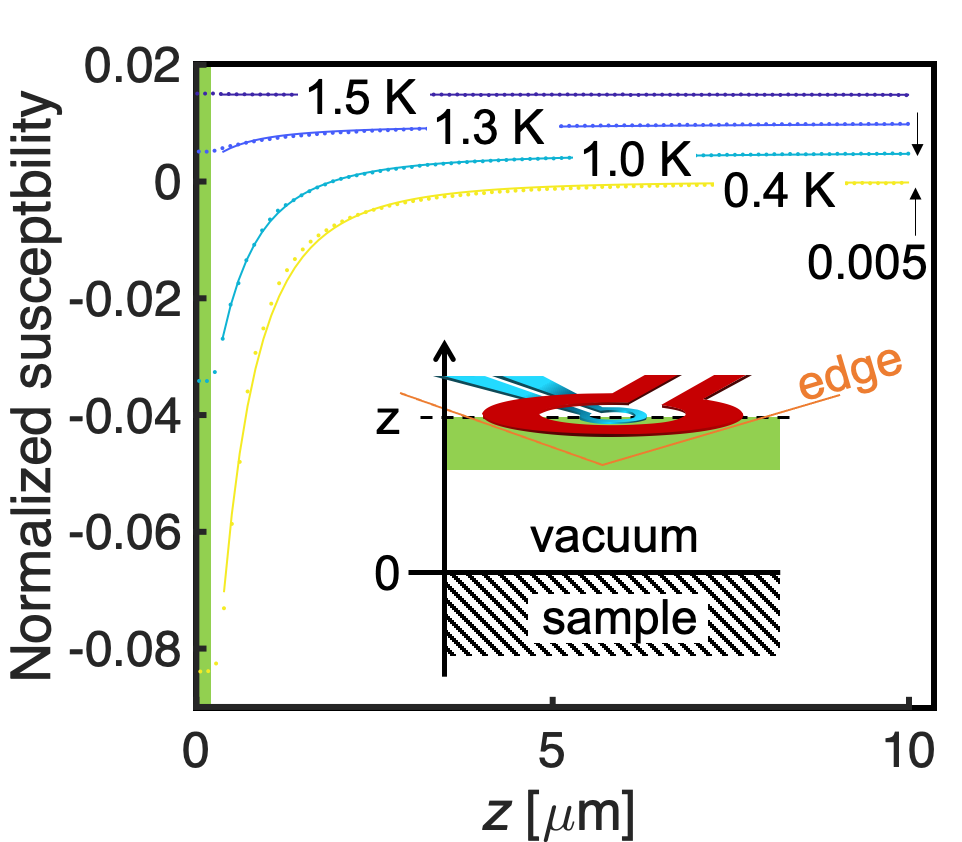}
			\caption{Penetration depth was calculated from the relation between the normalized susceptibility and the height of the pickup loop. Touchdown measurements of P2 at 1.5, 1.4, 1.0, and 0.35 K. Closed circles are experimentally obtained data. Solid lines are fitting curves. Inset shows the schematic of our experimental setup. The green area denotes the thickness of the shield layer.}
		\end{center}
	\end{figure}

    The electric current density of electrons and holes is given by
    \begin{equation}
        \mathbf{j} = -e^2\left(\frac{n_e}{m_e^*}+\frac{n_h}{m_h^*}\right) \mathbf{A}
    \end{equation}
    where we assume all electrons and holes form Cooper pairs and averaged canonical momentum is zero, and $n_e,n_h$ are the carrier densities of electrons and holes. Substituting $\mathbf{j}$ into the Maxwell equation, we obtain the London penetration depth by
    \begin{equation}
        \lambda = \sqrt{\frac{m_0}{\mu_0e^2}\frac{1}{n_em_0/m_e^*+n_hm_0/m_h^*}}.
    \end{equation}
    
    \begin{figure}[b]
		\begin{center}
			\includegraphics*[width=9cm]{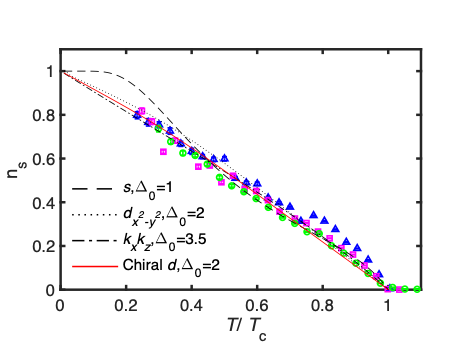}
			\caption{Simulated superfluid density temperature dependence from single band models did not capture details of temperature dependence of experimental data. ( The red squares, blue triangles, and green circles are measured data at P2, P3, and P4, respectively. All gap energies were represented in the unit of $k_{\mbox{\scriptsize B}}T_{\mbox{\scriptsize c}}$.}
		\end{center}
	\end{figure}

    The temperature dependence of the superfluid density in unconventional superconductors is estimated by the semi-classical approach with an anisotropic gap function $\Delta({\textbf k}, T) = g({\textbf k})\Delta_0(T)$, where $\Delta_0(T) = \Delta_0(0) \tanh{(\pi k_{\mbox{\scriptsize B}}T_{\mbox{\scriptsize c}}/\Delta_0(0)\sqrt{a(T_{\mbox{\scriptsize c}}/T-1)})} $ is obtained by solving the self-consistent gap equation with two free parameters, $\Delta_0(0) $ and $a$\cite{Prozorov2006}. We calculated the superfluid density for all {\it d}-wave symmetry gap functions in tetragonal symmetry\cite{sigristueda1991}. The chiral {\it d}-wave symmetry model $k_z(k_x+ik_y)$ was given by $g=2\sin{\theta}\cos{\theta}$, $k_xk_y$ model by $g=\sin{2\phi}$, $k_{x^2-y^2}$ model by $g=\sin{2\phi-\pi/4}$, $k_xk_z$ model by $g=\sin{\theta}\sin{\phi}$, and $k_z(k_x+k_y)$ by $g=\sin{\theta}\sin{\phi-\pi/4}$. The calculated superfluid density of single band {\it s}-wave and {\it d}-wave symmetries were shown in Supplemental Fig. S5.

    We also fitted the experimental data with the two band model with two free parameters $\Delta_{\mbox{\scriptsize h}}$ and $\Delta_{\mbox{\scriptsize e}}$, where $\Delta_0(T) = \Delta_i \tanh{(\pi k_{\mbox{\scriptsize B}}T_{\mbox{\scriptsize c}}/\Delta_0(0)\sqrt{a(T_{\mbox{\scriptsize c}}/T-1)})}$ and $i$ = e,h. The calculated curves for the model of chiral {\it d}-wave symmetry at both electron and hole bands well explain the experimental result at P2 with $\Delta_{\mbox{\scriptsize h}}=1.7 k_{\mbox{\scriptsize B}}T_{\mbox{\scriptsize c}}$ and $\Delta_{\mbox{\scriptsize e}}=5 k_{\mbox{\scriptsize B}}T_{\mbox{\scriptsize c}}$, and the results at P3 and P4 with $\Delta_{\mbox{\scriptsize h}}=1.85 k_{\mbox{\scriptsize B}}T_{\mbox{\scriptsize c}}$ and $\Delta_{\mbox{\scriptsize e}}=10 k_{\mbox{\scriptsize B}}T_{\mbox{\scriptsize c}}$ (Fig. 4c, Supplemental Fig. S6a). We also fitted the data with other two-band chiral {\it d}-wave models (such as $L_{\mbox{\scriptsize H}}+2P$ at the light hole band and $2P$ at the heavy electron band; Supplemental Figs. S6b-e), other {\it d}-wave symmetry models (such as $k_xk_z$ (or $k_z(k_x+k_y)$) and $k_x^2-k_y^2$ (or $k_xk_y$); Supplemental Figs. S7a,b), and a two-band isotropic {\it s}-wave model (Supplemental Fig. S7c).

	\begin{figure}[htb]
		\begin{center}
			\includegraphics*[width=16cm]{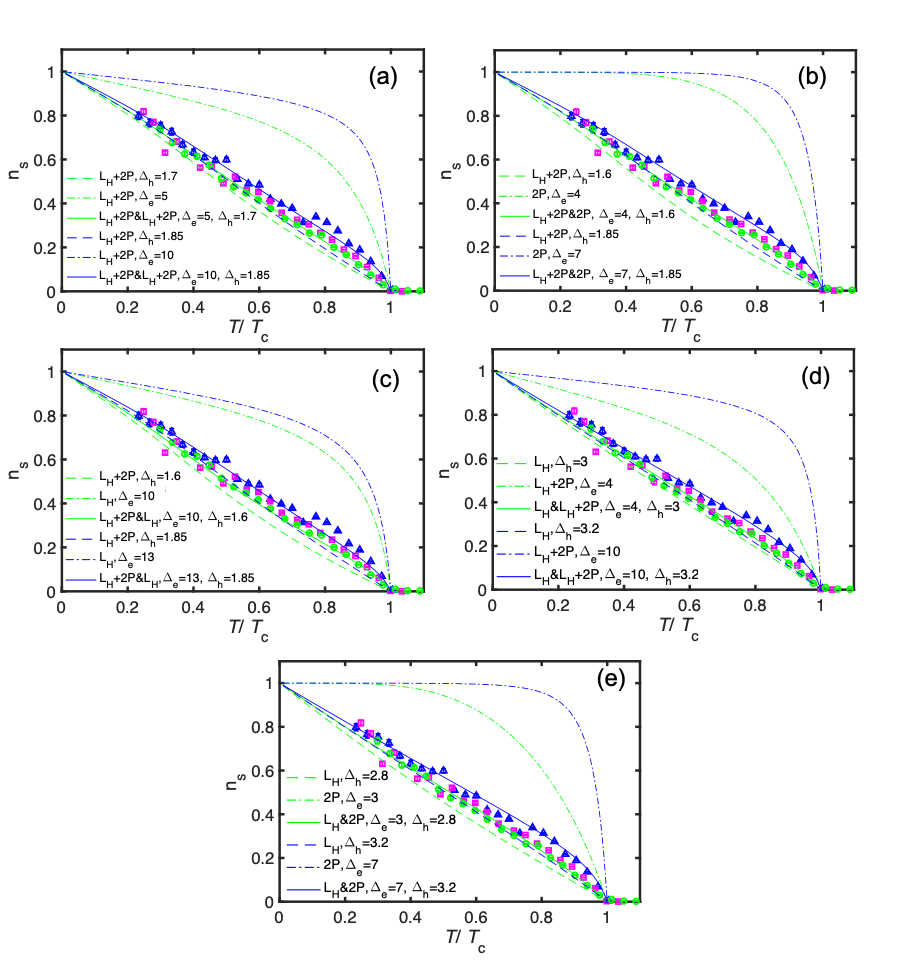}
			\caption{Two band gap models of chiral {\it d}-wave $k_z(k_x+ik_y)$ fitted well our experimental results of the temperature dependence of superfluid density. Solid lines are total superfluid densities of hole and electron bands. Dashed lines are superfluid densities of hole band. Dash-dotted lines are superfluid densities of electron band. (a) $L_{\mbox{\scriptsize H}}+2P$ (hole band) and $L_{\mbox{\scriptsize H}}+2P$ (electron band). (b) $L_{\mbox{\scriptsize H}}+2P$ (hole band) and $2P$ (electron band). (c) $L_{\mbox{\scriptsize H}}+2P$ (hole band) and $L_{\mbox{\scriptsize H}}$ (electron band). (d) $L_{\mbox{\scriptsize H}}$ (hole band) and $L_{\mbox{\scriptsize H}}+2P$ (electron band). (e) $L_{\mbox{\scriptsize H}}$ (hole band) and $2P$ (electron band). ( The red squares, blue triangles, and green circles are measured data at P2, P3, and P4, respectively. All gap energies were represented in the unit of $k_{\mbox{\scriptsize B}}T_{\mbox{\scriptsize c}}$.}
		\end{center}
	\end{figure}
	
	\begin{figure}[htb]
		\begin{center}
			\includegraphics*[width=16cm]{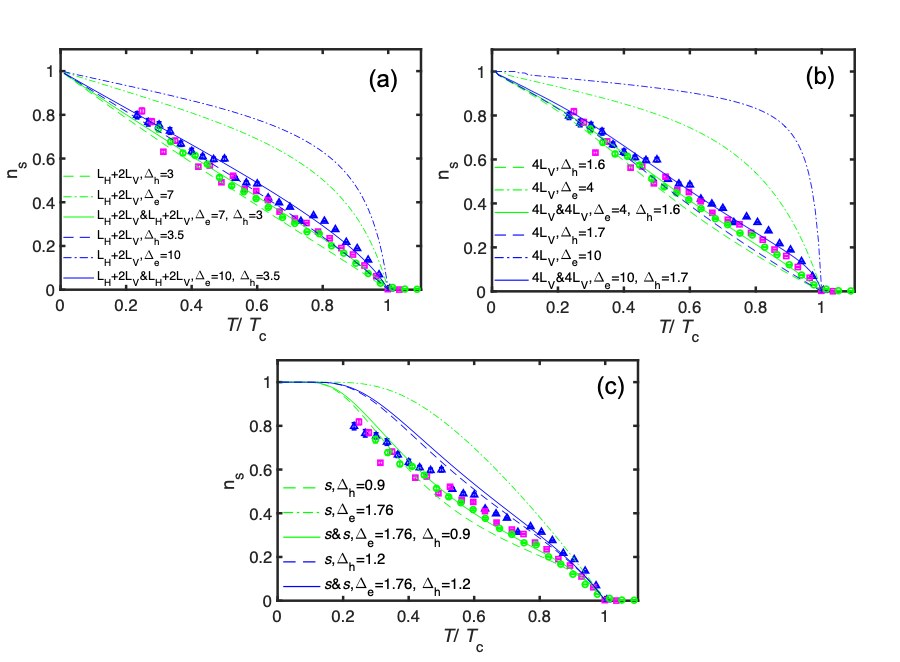}
			\caption{Two band gap models of time-reversal symmetric {\it d}-wave symmetries fitted better our experimental results of the temperature dependence of superfluid density than two band models of isotropic {\it s}-wave. (a)-(c) Simulated superfluid density temperature dependence from two-band gap models. Solid lines are total superfluid densities of hole and electron bands. Dashed lines are superfluid densities of hole band. Dash-dotted lines are superfluid densities of electron band. (a) Two band models of $k_xk_z$ or $k_z(k_x+k_y)$. $L_{\mbox{\scriptsize H}}+2L_{\mbox{\scriptsize V}}$ (hole band) and $L_{\mbox{\scriptsize H}}+2L_{\mbox{\scriptsize V}}$ (electron band). (b) Two band models of $k_x^2-k_y^2$ or $k_xk_y$. $4L_{\mbox{\scriptsize V}}$ (hole band) and $4L_{\mbox{\scriptsize V}}$ (electron band). (c) Two band models of isotropic {\it s} (hole band) and isotropic {\it s} (electron band). ( The red squares, blue triangles, and green circles are measured data at P2, P3, and P4, respectively. All gap energies were represented in the unit of $k_{\mbox{\scriptsize B}}T_{\mbox{\scriptsize c}}$.}
		\end{center}
	\end{figure}
	
	\begin{figure}[htb]
		\begin{center}
			\includegraphics*[width=16cm]{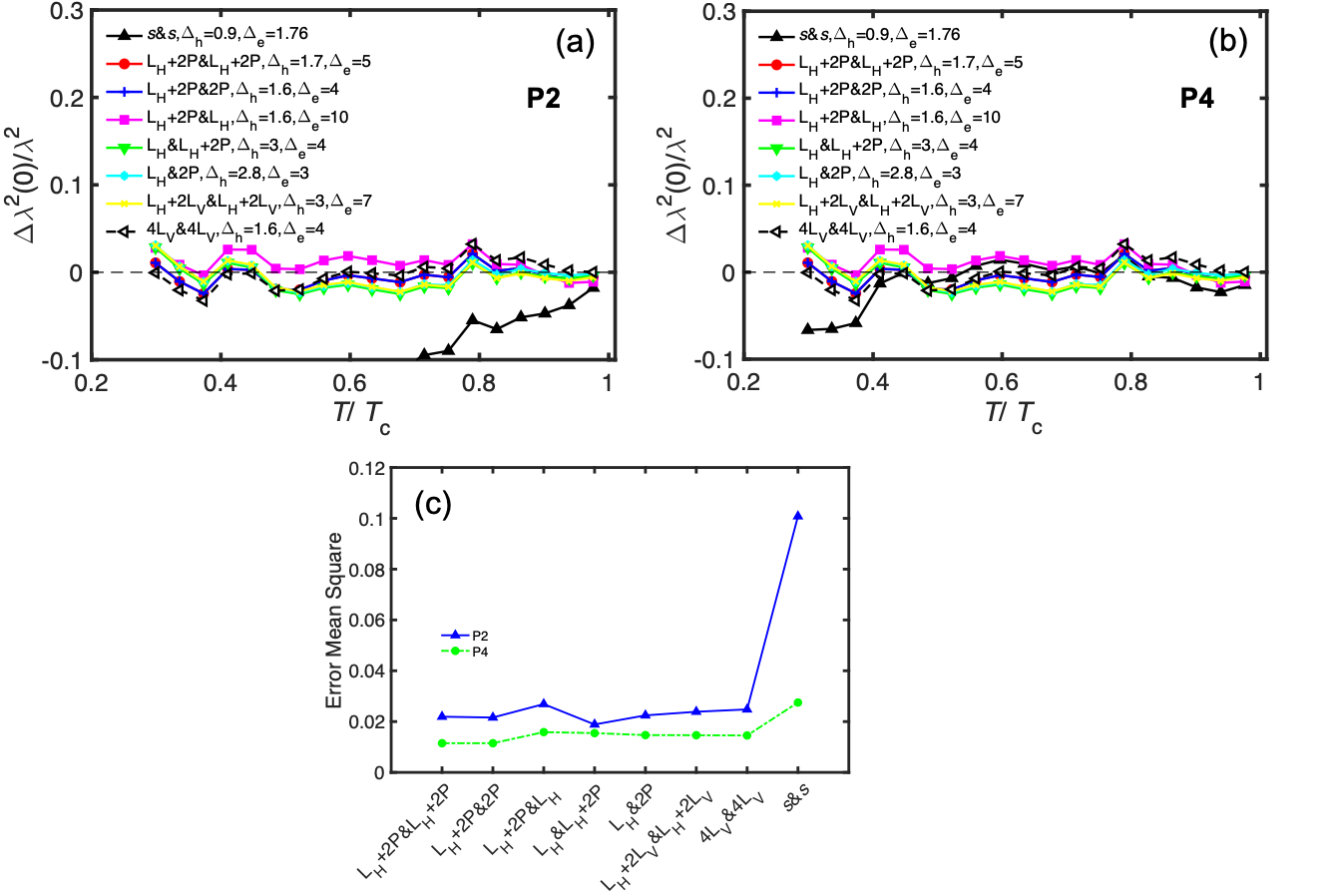}
			\caption{Fittings of {\it d}-wave models were better than that of isotropic {\it s}-wave models for two band gap models. (a),(b) Differences between simulated data and measured data at (b) P2 and (c) P4. All gap energies were represented in the unit of $k_{\mbox{\scriptsize B}}T_{\mbox{\scriptsize c}}$. (c) Error mean square of difference of simulated data and measured data at P2 and P4 for two-band gap models.}
		\end{center}
	\end{figure}


\begin{thebibliography}{}
		\bibitem{Mydosh2011} J. A. Mydosh and P. M. Oppeneer, Rev. Mod. Phys. {\bf 83}, 1301 (2011).
		\bibitem{Shibauchi2014} T. Shibauchi, H. Ikeda and Y. Matsuda, Philos. Mag. {\bf 94}, 3747 (2014).
		\bibitem{Broholm1987} C. Broholm, H. Lin, P. T. Matthews, T. E. Mason, W. J. L. Buyers, and M. F. Collins, Phys. Rev. B {\bf 43}, 12809 (1991). 
		\bibitem{Okazaki2011s} R. Okazaki, T. Shibauchi, H. J. Shi, Y. Haga, T. D. Matsuda, E. Yamamoto, Y. Onuki, H. Ikeda, and Y. Matsuda, Science {\bf 439}, 331 (2011).
		\bibitem{Tonegawa2012} S. Tonegawa {\it et al.}, Phys. Rev. Lett. {\bf 109}, 036401 (2012).
		\bibitem{Riggs2014} S. C. Riggs, M. C. Shapiro, A. V. Maharaj, S. Raghu, E. D. Bauer, R. E. Baumbach, P. Giraldo-Gallo, M. Wartenbe, and I. R. Fisher, Nat. Commun. {\bf 6}, 6425(2014).
		\bibitem{Tonegawa2014} S. Tonegawa {\it et al.}, Nat. Commun. {\bf 5}, 1(2014).
		\bibitem{Choi2018} J. Choi, O. Ivashko, N. Dennler, D. Aoki, K. von Arx,S. Gerber, O. Gutowski, M. H. Fischer, J. Strempfer, M. v. Zimmermann, and J. Chang, Phys. Rev. B {\bf 98}, 241113(R) (2018).
		\bibitem{Kiss2005} A. kiss and P. Fazekas, Phys. Rev. B, {\bf 71}, 054415 (2005).
		\bibitem{Haule2009} K. Haule and G. Kotliar, Nat. Phys. {\bf 5}, 796 (2009).
		\bibitem{Kusunose2011} H. Kusunose and H. Harima, J. Phys. Soc. Jpn. {\bf 80}, 084702 (2011).
		\bibitem{Cricchio2009} F. Cricchio, F. Bultmark, O. Gr\aa{}n\"{a}s, and L. Nordstr\"{o}m, Phys. Rev. Lett. {\bf 103}, 107202 (2009).
		\bibitem{Ikeda2012} H. Ikeda, M.-T. Suzuki, R. Arita, T. Takimoto., T. Shibauchi, and Y. Matsuda, Nat. Phys.  {\bf 8}, 528 (2012).
		\bibitem{Knetsch1993} E. A. Knetsch, A. A. Menovsky, G. J. Nieuwenhuys, J. A. Mydosh, A. Amato, R. Feyerherm, F. N. Gygax, A. Schenck, R. H. Heffner, and D. E. MacLaughlin, Physica B {\bf 186-188}, 300 (1993).
		\bibitem{Hattori2018} T. Hattori, H. Sakai, Y. Tokunaga, S. Kambe, T. D. Matsuda, and Y. Haga, Phys. Rev. Lett. {\bf 120}, 027001 (2018).
		\bibitem{Brison1995} J. P. Brison, N. Keller, A. Vernière, P. Lejay, L. Schmidt, A. Buzdin, J. Flouquet, S. R. Julian, and G. G. Lonzarich, Physica C {\bf 250}, 128 (1995).
		\bibitem{Hasselbach1993_46} K. Hasselbach, J. R. Kirtley, P. Lejay, Phys. Rev. B {\bf46} 5826 (1993).
		\bibitem{Hasselbach1993_47} K. Hasselbach, J. R. Kirtley, and J. Flouquet, Phys. Rev. B {\bf47} 509 (1993).
		\bibitem{Yano2008} K. Yano,	T. Sakakibara, T. Tayama, M. Yokoyama, H.  Amitsuka, Y. Homma, P. Miranović, M. Ichioka, Y. Tsutsumi, K. Machida, Phys. Rev. Lett. {\bf 100}, 017004 (2008). 
		\bibitem{Kittaka2016} S. Kittaka, Y. Shimizu, T. Sakakibara, Y. Haga, E. Yamamoto, Y. Ōnuki, Y. Tsutsumi, T. Nomoto, H. Ikeda, and K. Machida, J. Phys. Soc. Jpn. {\bf 85}, 033704 (2016).
		\bibitem{Kohori1996} Y. Kohori, K. Matsuda, and T. Kohara, J. Phys. Soc. Jpn. {\bf 65} 1083 (1996).
		\bibitem{Kasahara2007} Y. Kasahara, T. Iwasawa, H. Shishido, T. Shibauchi, K. Behnia, Y. Haga, T. D. Matsuda, Y. Onuki, M. Sigrist, and Y. Matsuda, Phys. Rev. Lett. {\bf 99}, 116402 (2007). 
		\bibitem{Kasahara2009} Y. Kasahara, T. Iwasawa, H. Shishido, T. Shibauchi, K. Behnia, T. D. Matsuda, Y. Haga, Y. Onuki, M. Sigrist, and Y. Matsuda, J. Phys.: Conf. Ser. {\bf 150}, 052098 (2009).
		\bibitem{Yamashita2015} T. Yamashita {\it et al.}, Nat. Phys. {\bf 11}, 17 (2015).
		\bibitem{Schemm2015} E. R. Schemm, R. E. Baumbach, P. H. Tobash, F. Ronning, E. D. Bauer, and A. Kapitulnik, Phys. Rev. B {\bf 91}, 140506(R) (2015).
		%\bibitem{Sumiyoshi2014} H. Sumiyoshi and S. Fujimoto, Phys. Rev. B {\bf 90}, 184518 (2014).
		\bibitem{Uemura2005} S. Uemura, G. Motoyama, Y. Oda, T. Nishioka, and N. K. Sato, J. Phys. Soc. Jpn. {\bf 74}, 2667 (2005).
		\bibitem{Amitsuka2007} H. Amitsuka, K. Matsuda, I. Kawasaki, K. Tenya, M. Yokoyama, C. Sekine, N. Tateiwa, T. C. Kobayashi, S. Kawarazaki, and H. Yoshizawa, J. Magn. Magn. Mater. {\bf 310}, 214 (2007).
		%\bibitem{Kawasaki2014} I. Kawasaki, I. Watanabe, A. Hillier, and D. Aoki, J. Phys. Soc. Jpn. {\bf 83}, 094720 (2014).
		%\bibitem{chiralsc} C. Kallin and J. Berlinsky, Chiral superconductors, Rep. Prog. Phys. {\bf 79} 054502 (2016). 
		%\bibitem{maenoreview} A. P. Mackenzie and Y. Maeno, Rev. Mod. Phys. {\bf 75}, 657(2003).
		\bibitem{Cliff2010} C. W. Hicks, J. R. Kirtley, T. M. Lippman, N. C. Koshnick, M. E. Huber, Y. Maeno, W. M. Yuhasz, M. B. Maple, and K. A. Moler, Phys. Rev. B {\bf 81}, 214501 (2010). 
		\bibitem{Hykel2014} D. J. Hykel, C. Paulsen, D. Aoki, J. R. Kirtley, and K. Hasselbach, Appl. Phys. Rev. B {\bf 90}, 184501 (2014).
		%\bibitem{Kasai2018} J. Kasai, Y. Okamoto, K. Nishioka, T. Takagi, and Y. Sasaki, Phys. Rev. Lett. {\bf 120}, 205301 (2018).
		\bibitem{Irene} I. P. Zhang,  J. C. Palmstrom, H. Noad, L. B.-V. Horn, Y. Iguchi, Z. Cui, E. Mueller, J. R. Kirtley, I. R. Fisher, and K. A. Moler, Phys. Rev. B {\bf 100}, 024514 (2019).
		\bibitem{Beena2010} B. Kalisky,  J. R. Kirtley, J. G. Analytis, Jiun-Haw Chu, A. Vailionis, I. R. Fisher, and K. A. Moler, Phys. Rev. B {\bf 81}, 184513 (2010).
		\bibitem{Logan2019} L. B.-V. Horn, Z. Cui, J. R. Kirtley, and K. A. Moler, Rev. Sci. Instrum. {\bf 90}, 063705 (2019).
		\bibitem{kirtleyprb2012} J. R. Kirtley {\it et al.}, Phys. Rev. B {\bf 85}, 224518 (2012).
		\bibitem{supple} See Supplemental Material at [URL will be inserted by publisher] for (Sec.1) the details of experimental setup,  the details of estimation of chiral domain fields, the relation between the superconductivity and ferromagnetic domains in sample 1 and 2, (Sec.2) the details of isolated vortex dynamics simulation and the detailed discussion of the origin of anisotropic pinning force, and (Sec.3) the detailed simulation of superfluid density.
		\bibitem{kirtleyrsi2016} J. R. Kirtley {\it et al.}, Rev. Sci. Instrum. {\bf 87}, 093702 (2016).
		\bibitem{Ishikawa1977} M. Ishikawa, Prog. Theor. Phys. {\bf 57} 1836 (1977).
		\bibitem{Tada2015} Y. Tada, W. Nie and M. Oshikawa, Phys. Rev. Lett. {\bf 114}, 195301 (2015).
		\bibitem{Nie2020} W. Nie, W. Hwang and H. Yao, Phys. Rev. B {\bf 102}, 054502 (2020).
		\bibitem{Tonegawa2013} S. Tonegawa {\it et al.}, Phys. Rev. B {\bf 88}, 245131 (2013).
		\bibitem{Maple1986} M. B. Maple, J. W. Chen, Y. Dalichaouch, T. Kohara, C. Rossel, M. S. Torikachvili, M. W. McElfresh, and J. D. Thompson, Phys. Rev. Lett. {\bf 56}, 185 (1986).
		\bibitem{Ohkuni1999} H. Ohkuni et al., Philos. Mag. B {\bf 79}, 1045 (1999).
		\bibitem{Okazaki2008} R. Okazaki, Y. Kasahara, H. Shishido, M. Konczykowski, K. Behnia, Y. Haga, T. D. Matsuda, Y. Onuki, T. Shibauchi, and Y. Matsuda, hys. Rev. Lett. {\bf 100}, 037004 (2008).
		\bibitem{MatsumotoSigrist1999} M. Matsumoto and M. Sigrist, J. Phys. Soc. Jpn. {\bf 68}, 994(1999).
		\bibitem{Bluhm2007} H. Bluhm, Phys. Rev. B {\bf 76}, 144507 (2007).
		\bibitem{Matsuda2001} K. Matsuda, Y. Kohori, T. Kohara, K. Kuwahara, and H. Amitsuka, Phys. Rev. Lett. {\bf 87}, 087203(2001).
		\bibitem{Yokoyama2005} M. Yokoyama, H. Amitsuka, K. Tenya, K. Watanabe, S. Kawarazaki, H. Yoshizawa, and J. A. Mydosh, Phys. Rev. B {\bf 72}, 214419(2005).
		%\bibitem{Campbell1972} A.M. Campbell and J.E. Evetts, Adv. Phys. {\bf 21}, 199(1972).
		\bibitem{Ullmaier1975} H. Ullmaier, Irreversible Properties of Type II superconductivity, (Berlin: Springer, 1975) pp42-43
		\bibitem{Golosovsky1996} M. Golosovsky, M. Tsindlekht and D. Davidov, Supercond. Sci. Technol. {\bf 9}, 1(1996).
		\bibitem{Amato1997} A. Amato, Rev. Mod. Phys. {\bf 69}, 1119 (1997).
		\bibitem{sigristueda1991} M. Sigrist and K. Ueda, Rev. Mod. Phys. {\bf 63}, 239 (1991).
		\bibitem{Prozorov2006} R. Prozorov and R. W. Giannetta, Supercond. Sci. Technol. {\bf 19}, R41 (2006).
		\bibitem{Okazaki2011} R. Okazaki {\it et al.}, J. Phys.: Conf. Ser. {\bf 273}, 012081 (2011).
		


	\end{thebibliography}
\end{document}